# SECURE DIGITAL ADMINISTRATION IN MEDICAL ENVIRONMENT

Diana Berbecaru, Antonio Lioy, Marius Marian
*Dip. di Automatica e Informatica, Politecnico di Torino, Torino, ITALY*

Diana Marano
*I&T - Informatica e Telecomunicazioni S.p.A Roma, ITALY*

**ABSTRACT**

The efficiency and service quality in a medical environment can be improved by using electronic documents (or e-docs) and digital signatures to speed up both doctors' activity and to provide in the same time easy retrieval and use of needed data without loosing convenience. Our proposed solution satisfies these needs by making use of the AIDA system and many cutting-edge techniques to build a digitalized management system. To be more specific, we present firstly the AIDA document exchange framework for the management (creation, search, storage) of e-docs expressed in XML format. The framework provides security services, like privacy, authorization, authentication and non-repudiation. We describe next the use of AIDA in a real world medical service, namely the creation of electronic medical records (e-MRs). Doctors can use mobile devices that embed a secure trustworthy environment defined also in AIDA for digitally signing the e-MRs. Technically speaking a handheld PC equipped with a smart-card reader and integrating What You See Is What You Sign (WYSIWYS) features will be used for viewing and signing the e-MRs. Furthermore the proposed system is easily integrated with the infrastructure (e.g., database system) already in use at hospital's administration site and allows easy handling and updating of data processed on the mobile devices. The use of web interface for the operations to be executed on the mobile device or for those executed on the remote part of the system makes the whole application homogeneous and easy to use.

## 1. INTRODUCTION

Despite the growth of Internet and the appearance of new technologies in the field of digital signatures, the implementation of secure digital services is still an open issue. For example the traditional medical processes continue to be based on paper documents. The paper-based systems usually present several disadvantages due to large amounts of papers handled and delays in documents' processing and delivery. This paper describes the implementation of a secure digital solution for the creation of medical records. The solution makes use of the Advanced Interactive Digital Administration (AIDA) system [1]. Doctors use mobile terminals as instruments for preparing for visits, writing diagnoses, and signing the e-MRs produced. The main objective in designing our solution was to offer certain key characteristics, particularly of interest for the end-users, such as:

*Usefulness and transparency.* The solution must allow doctors to work either on-line or off-line, in isolated environments or connected to a remote server without asking them to know the technical details.
*Flexibility.* The solution can be easily adapted if the workflow changes in time or for use in other medical environments, like divisions, hospitals or private clinics. Various workflows should be easily modeled when the e-docs' content or the internal management of e-docs changes.
*Security.* The security features must be maintained when a completely digitalized system is used. Security was one of the major requirements taken into consideration also at the AIDA system's design.

The medical environment is a typical environment where a large amount of sensitive information is handled and where the users are operating mostly off-line. Small personal devices, like notebooks or PDAs that are equipped with a smart-card reader and that integrate WYSIWYS features, seem to be the ideal solution for secure signing equipment: the devices are small and can be carried around easily, while the display is usually still capable of displaying different forms of data to be signed [6]. With the increasing number of persons using mobile devices (e.g., PDAs, handheld PCs), the challenge is to improve the functionality of these devices by allowing users to securely process (view, sign and store) the e-docs off-line or send them to the communicating parties in real time. The AIDA system gives a solution for this category of users having weak computing capabilities and operating mostly off-line as it is described in the present paper. The paper is organized as follows: Section 2 describes the components of the AIDA system and its functionality, Section 3 describes the tools used for data management in AIDA, Section 4 provides details on the 'off-line' solution used in the medical environment and Section 5 summarizes the conclusions.

## 2. AIDA COMPONENTS AND OPERATION

The AIDA system, developed inside the homonymous EC-funded project, implements a secure technical environment for XML e-doc management that can be easily personalized by the e-administration service providers (ESP) for different administration scenarios from universities, hospitals or companies. The system gives a solution to the creation of a trustworthy environment for the generation of digital signatures and as well as to express paper data in an equivalent machine-readable format. The AIDA's e-doc management platform allows the automatic generation of e-docs having a predefined content and making use of data inserted by the user at run-time. The platform includes a secure infrastructure that support the whole life cycle of an e-doc and a terminal used to display the e-docs in a way that impedes security attacks [6] at digital signing time or when verifying the e-docs' content and the signature(s) applied on them. The security features in AIDA have been widely described in [3]. XML is appropriate to be used for the e-doc's format in order to avoid security attacks [10]. Smart-cards [11, 12] are used for authentication and signing purposes, the securing of the communication channels is done by employing the SSL protocol and X.509 certificates are used also for access control. Also the trustworthy creation/viewing of the signatures using WYSIWYS software has been carefully addressed. The system can be easily incorporated to make use of the back-end (e.g., a database system) already existent and in use at the administration side. Thus, the public bodies should not have to modify their currently used Database Management System when adding support for AIDA. The system allows also for the existent front-ends (PCs or other ATM-like devices) to be also used, allowing users to perform simple electronic transactions [3].

AIDA framework conforms to a web application architecture that typically follows an $n$-tier model [2]. The presentation level includes not only the web browser but also the web server, which is responsible for assembling the data into a presentable format. At this level there are included the HTML forms, the Java applets or any other system capable of presenting data like the WYSIWYS module. The application level consists basically of the code, which the user calls upon through the presentation layer to retrieve the desired data. It is further the task of the presentation logic to receive the data and format it for display. The data level contains the data that is needed for the application and could consist of any source of information. The data source is not limited to just an enterprise database like Oracle but could include also a set of XML [4] documents or a remote system. In this architecture the ESP database backend will work together with AIDA server (or AIDA platform - *Aplatform*) through interfaces implemented at application layer to supply data to the presentation layer. In AIDA the presentation layer consists of HTML forms and standalone client applications that incorporate WYSIWYS features. The presentation and the application levels are incorporated at the ESP and user levels.

The client-server paradigm is used to implement the overall system. This architecture also takes the form of a server-to-server configuration that is one server plays the role of a client and request services (e.g., database service) from another server. The *Aplatform* communicates via AIDA commands (*Acommands*) with client applications using an application protocol named *AIDA protocol* (or *Aprotocol*). Every *Aplatform* operation is explicitly requested by the Scenario Application (SA), which implements the actual service workflow. The SA is responsible for interaction with the Web server that will surely be present in any scenario to provide the user with an appropriate web portal. It gathers also proprietary data through appropriate interfaces and communicates with the *Aplatform* for specific functions like e-doc creation, display and verification. The *Aplatform* is located inside the ESP's intranet and is installed and configured using the administration console by the e-administrator.

The ESP's applications, called SAs, implement the workflows for various e-doc services. For this purpose the SAs interact with the *Aplatform* to create/store e-docs, with the user interfaces like a web browser, WYSIWYS client or with the database server. The *Aplatform* must not be confused with the web server that interacts with the users and runs SAs that in turn communicate with the inside *Aplatform*. The Application Server (AS), acting as a web server, is the connection point for service users while the *Aplatform* is never visible externally. The *Aplatform* includes the *definitions repository* that contains specifications and meta-information for the creation and verification of XML docs, like DTDs and XSL stylesheets and other internal data. The *Aplatform* includes also the *document directory* where are stored the signed e-docs. Currently the *Aplatform* has also a GUI that allows the administrator to configure the ports and the log files. Two equivalent application ports with similar functionality, called the *scenario* and *service ports*, are used to access the *Aplatform*. These ports can be configured to allow an ESP to freely choose the openness of the system that is they are used to exchange messages directly or via gateway programs. The ports can be set up either to specify the set (full or restricted) of *Acommands* accepted by the port or to indicate the outside visibility of the port. The gateway programs tunnel commands from one port to another one and connects clients situated outside the ESP's intranet to the *Aplatform*. This permits the SA to be situated anywhere in the Internet or in the ESP' intranet. The *administration port* is used to start/stop the other ports and to accept administration functions. At start-up the AIDA server requires the administrator to set-up a number of parameters like the role set certificate (technically this is an X.509 certificate used by the role administrator to set up other roles as described in Section 3), the ODBC parameters (e.g., driver name) used to connect to the proprietary database or the *Acommands* allowed to be received on the service and scenario ports. The *Aprotocol* uses signed XML structures for the AIDA messages exchanged. A special Document Type Definition (DTD) is used for the syntax of these messages. AIDA messages can be sent directly to the *Aplatform* using the *Aprotocol* or can be tunneled via HTTP to the gateways that send them to the *Aplatform* on dedicated ports. The Security Service Bricks (SSB) are the modules that support the security service and enable most of AIDA' secure functionality like support for smart cards, SSL or XML signatures [5]. The SSB are never used directly by the SA developers because they are incorporated into the upper-level packages and tools of the system (Section 3). The SA developers will use an AIDA API for managing e-docs.

## 3. SOFTWARE TOOLS FOR DATA MANAGEMENT IN AIDA

**Definitions Manager (DefMan)** is a client tool used at the AIDA installation to define the types of e-docs used. The actual e-docs' content is defined at this time. As the format of e-docs is XML then at this step the elements and attributes are specified (XML Schemas) as well as the transformations used for displaying the e-doc's content (XSL stylesheets) and the name of the fields in the HTML form for the user's web interface. The SA that implements the workflow for that service will use the e-doc types specified for the service. DTDs and XSL stylesheets are also XML structures and are signed with the *definer* role. They are afterwards embedded into *Acommands* sent towards the *Aplatform*. The *Aplatform* allows installation of the schemas and of the transformations only if the definer role has been previously set up with the *RoleMan*.

**Scenario Desktop (ScenDesk)** is a Graphical User Interface (GUI) client tool used for interactive operations and handling of e-docs. Persons called *referents* that are responsible for certain tasks in a public service (like processing a request) use this tool to model the operations in the ESP's workflows that need human processing. With *ScenDesk* the referents can view, create, store, sign and search for specific e-docs stored on the *Aplatform*. To modify e-doc' status (e.g., from *pending* to *processed*) the e-doc *attributes* are defined. They allow the *Aplatform* to fulfill a variety of functions and support various models of e-doc structures, meta-information and various ESP e-doc workflows. No single system could be built to incorporate every aspect of e-doc management preferred by an ESP and his existent business logic. To handle *Aplatform*'s data stores as needed (e.g., to partition the document directory into *input* and *output* e-doc parts), to add specific data to e-docs (e.g., *state*) or to configure type-specific data need by the SA (e.g., to mark some fields as optional) the so-called *attributes* are used to hold information needed by the ESP. Unlike *static* attributes, which are set once at the *Aplatform* start-up and never changed later, there exist also *dynamic* attributes because the *ScenDesk* can read and set them. Referents can create *output* e-docs using field data extracted from one or more existing XML e-docs serving as input source (called *input* e-docs) as well as field data entered directly via the GUI. The connection between the *input* and *output* e-docs is not represented in the definitions repository but it is specified using *processing rules* inserted into an XML *configuration file* stored on the referent PC. Fields in the *output* e-doc not automatically filled by the *ScenDesk* after applying the *processing rules* are presented to the referent in a dialog box and their values are entered manually.

**Role Manager (RoleMan)**. The role administrator uses this GUI client tool to define and manage various roles in AIDA by installing the role certificates and setting up afterwards restrictions on the *Acommand*s and e-doc types each referent is allowed to generate and view. The communication between the *Aplatform* and any client module (WYSIWYS client, *ScenDesk*, *DefMan*) is done via signed *Acommand*s. In practice, the client modules sign every *Acommand* sent to the *Aplatform* with a private key and attach the role certificate containing the pair public key. The *Aplatform* checks if the role that requested the execution of the *Acommand* has the privileges to do so by looking at the *role map* that connects each role with the *Acommand*s and the document types that are allowed for it. Technically speaking, this is done by comparing the certificate fingerprint of the role certificate sent together with the *Acommand*'s data against the fingerprints stored within the *role map*. If no match is found, then this means that the role is unknown and the requested *Acommand* is rejected. If a match exists then the *Aplatform* will execute the *Acommand*.

**YSIWYS software** is either a standalone application running on the client's terminal (PC or mobile device) or is embedded in the other client tools, e.g., *ScenDesk*. It satisfies some rough requirements like viewing, signing or verifying the e-docs. It incorporates a XML viewer for displaying XML docs, a library for signing XML docs and a cryptographic library for digital signing that makes use of smart cards to generate signatures [7][8][9]. By using a set of signed *Acommands* the WYSIWYS client asks the *Aplatform* execute a number of tasks, like for example to view an XML e-doc by applying transformation rules on the content according to the user's choice (e.g., language). Each *Acommand* request is signed with a role certificate, which is built in into the viewer of the WYSIWYS client together with the corresponding private key and will be the same for all instances of the program. In this sense, a request from the WYSIWYS standalone client is 'anonymous' because every user using an instance of the viewer uses the same role certificate. This generic WYSIWYS role certificate must be installed also in the role map of the *Aplatform* using the *RoleMan*. Most *Acommand*s have e-docs that they operate on either explicitly or implicitly. Consequently, besides allowing only certain *Acommand*s to be executed by a specific role, the *Acommand*'s execution could be further restricted by defining a subset of e-doc types it can operate on. This mechanism will be useful for *ScenDesk*'s functionality because different roles can be created for different referents using the same *Acommand*s but each referent will be able to work on the e-docs of specific types he is responsible for. The role certificates allow access control by restricting the user's actions to the ones his ESP sets up according to his internal responsibility scheme. The *ScenDesk* and/or *DefMan* interact with the *Aplatform* either directly via the *scenario port* or indirectly via gateway programs that forwards requests to the *scenario port*. The latter case is very useful and necessary when the *ScenDesk* is run outside ESP's intranet. The communication between the *ScenDesk* and the *Aplatform* is completely transparent because the application support library handles it. This library provides a number of classes and methods that resemble functionality which is either directly executed on the client' side or implicitly involves communication with the *Aplatform* using the *Aprotocol*. The WYSIWYS client communicates instead with the *Aplatform* either directly through the service port or indirectly via the *service port* gateway program that forwards the requests to the *service port*.

## 4. A CASE STUDY

This section describes the use of the AIDA system to implement a solution for the creation of e-MRs inside the Angiology division of S. Camillo - Forlanini Hospital in Rome (Italy), which agreed to participate in our trial. This solution was developed in order to increase the quality of the actual service, to reduce the patient's wait time to get medical reports or to allow persons to safely share sensitive information like the medical history of a patient. Our electronic solution allows physicians to work easily also in environments where even when it is necessary to access data stored in a remote server, it's impossible to use network connections. Also the given solution is not dedicated only to one particular scenario but it can be easily applied in different environments.

## 4.1 Current Organization of Medical Reports Release Service

The Angiology division contains a surgical department providing services related to instrumental examinations and medical checks both to *external* and *internal* patients. The first ones are persons that go to this surgery to undergo themselves to visits or specific examinations (e.g., Doppler) on their physician's request; the second ones are patients in other divisions of the hospital requiring visits or particular instrumental examinations required by the physicians of the original division. In both cases any type of service has to be reserved previously with a particular specialist.

For the established day all the booked visits are recorded in a list of patients, which is communicated to the specialist who has care of them. During the medical examination each doctor writes down or dictates his report with a dictaphone or to a secretary. Subsequently the report is written on a paper module checked and signed in the end by the doctor. The module is then enclosed with the *internal* patient's case sheet or is retrieved by the *external* patient that can use it as needed. In the traditional paper-based workflow we can note that the medical reports are manually compiled and consequently they are not immediately available, the risk of errors during the reports collection is remarkable, the patient's case sheet is a set of documents not easy manageable, which is conserved at least for ten years. We note also that only the patient can resume his own medical history, due to the absence of an objective memory and the medical reports can be produced in a single copy only.

## 4.2 Electronic Medical Reports Release Service

AIDA draws attention to itself by observing that the advantages for the public users are clear. With the AIDA-based design several requirements are met, like the accessibility, knowledge, security and usability of data, and data distribution. The AIDA system provides these features. Furthermore the end-user's convenience, expressed by the usability and simplicity of use by the physicians is achieved. This goal is met by designing and implementing a distributed application which renders to the physicians the possibility of using the system homogeneously and transparently in connection with a remote data server or in situations in which technical needs impose to work off-line. The proposed solution guarantees also hardware independence and use based on security services embedded in the AIDA system: in this way at the initialization phase the data is personalized with respect to users and each physician will be able to use an available mobile device and will be enabled to access precisely the e-docs he is in charge of.

The AIDA-based service requires setting up a minimal hardware configuration with one PC server, and some mobile devices used as signing terminals. The mobile device used was a HP Jornada 720. The connection between PC and mobile terminal can be made via serial cable o modem but the best choice we found for our purpose is the access via Wireless LAN, because it does not force the specialists to use a cable and offers greater mobility. In order to participate in services, every specialist has a handheld PC and a personal smart card. It is also possible to share one device among several persons. On the HPJornada720 it is installed the built-in WYSIWYS terminal sign application (called AIDASign) that allows the specialist to view, modify e-doc's field values and sign the e-docs. This application logically is part of the AIDA even though is doesn't have the whole functionality of the WYSIWYS client run on a PC. The signing certificate is stored on the smart card together with the corresponding private key. A role is defined for each specialist to restrict his access to various types of e-docs. The role is expressed by means of a role certificate i.e., an X.509 certificate whose task is to allow each specialist in the surgery to manage (create, store) *e-MR*s for the visit he made. A further improvement of the AIDA should allow the definition of groups of users in order to avoid the role administrator set up and handle a lot of roles. If groups are supported by the AIDA system then the addition/deletion of the users from the group could be done independently of AIDA, i.e., without having to modify the configuration of roles on the *Aplatform*. The Application Server (AS) runs MS Windows 2000 OS and has a Firewall, a Microsoft IIS, a TomCat servlet engine configured. The system makes use also of a MySQL database, the server component of the µ-Web architecture and the application depicted above to manage the surgical process. The Scenario Application (SA) is implemented as a set of Java servlets running on the AS. They communicate on one side with the *Aplatform* and with the database server and on the other side with the user as a typical web application. Production of electronic medical reports (*e-MR*) gives higher flexibility and usability to the medical services. Thus the automation allows redesigning the workflow as follows. Through the request, the *external* and *internal* patients can have his previously reserved visit or examination. During the visit, the specialist inserts the diagnosis into his own handheld PC, saves the e-MR locally and archives it also in a central repository within a form of electronic patient's "case sheet". Data from the e-MR is finally printed and the doctor applies a handwritten signature on the paper medical report to ensure its legal value.

*About performances.* For the personal and business productivity in the hospital, the visit reservation and registration services allow to manage more efficiently the medical work. The electronic registration of visits allows the specialist to quickly access some additional information about a particular patient. This data would be useful to the specialist for writing the actual diagnosis or to consult on line other possible results of previous medical examinations. When writing the medical report the specialist is interested in the diagnosis, while additional information such as patient's particulars, type of examination or date, are managed directly by the application. The reports for surgical visit and instrumental examination would be allowed to be stored also in an improved service, while presently only the reports for the patient admitted to hospital are stored. Surgical administrators can provide all the parameters needed for business reporting: number of examination, type of service, statistical data about overall surgical productivity, and instrumental and medical methodology. The physician can modify easily the diagnoses prior to the generation/signing of e-MR. The possibility of forging reports and medical certificates is reduced, while present systems are quite vulnerable. Medical history of patient is stored into the system and can be useful for other checks. Using a form with a predefined content and fields partially filled in allows uniformity in executing medical and instrumental examinations in accordance with international guidelines. Automation of the whole process aids also the specialist to avoid some unintentional mistakes in compiling forms.

## 4.3 Implementation Aspects

The implementation of a real service based on AIDA involves the interconnection of several components, some of which are not part of the *Aplatform* itself. This holds especially for the actual workflow logic that connects to proprietary resources to query user data or other information necessary to complete e-doc processing. Besides AS configuration, we used a particular software architecture developed by I&T named μ-Web architecture (Figure 1).

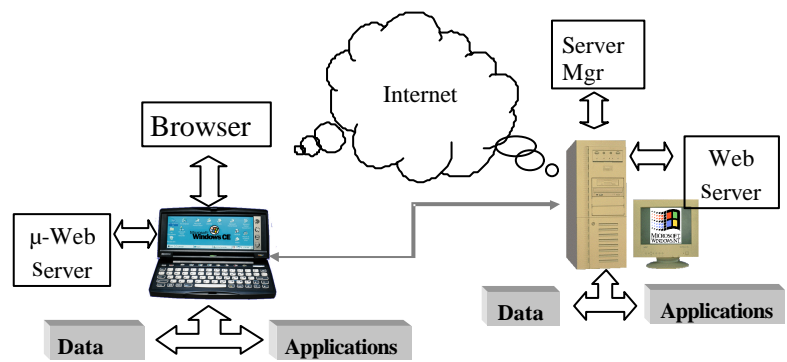

Figure 1. μ-web Architecture

The use of μ-Web is determined by the intrinsic off-line nature of the service described. However for other off-line simpler scenarios, architectures not incorporating μ-Web can be used if other mechanisms are used to synchronize data on the mobile device and the server. μ-Web has been developed to allow a web application execute on the handheld PC by using a local web server able to access data both from a remote data source as well as a database stored on the mobile device. In the same time μ-Web guarantees the independence of applications from a specific mobile device by using a web browser interface and allows clients to use mobile applications both either online or offline because the web server running on the mobile device can establish connections with a remote server on request. Finally μ-Web supports the porting of mobile applications on new devices in a simple, fast and transparent way. The μ-Web architecture comprises the components on the mobile device (Windows CE) like the Web server, Microsoft ADOCE, a web browser and the components on the remote server (Windows NTx) like the *ServerMgr*, RDBMS, ODBC Manager and optionally a web server. Besides the standard configuration that foresees the on-line connection to a remote web server by means of a network or dialup connection, another possible configuration allows to use the μ-Web's local web server to respond to the browser requests. Without a network connection, on the device run local web applications accessing only the device data and files stored locally. However, if an application tries to access data not present locally, the μ-web is able to connect to a remote server and download it.

Additionally, it is possible to have any other hybrid configuration where the μ–Web's API allows distribution of the web applications' tasks between the device and a remote server, if necessary, by accessing directly a remote data source or downloading data from the remote data source to the local store. The *ServerMgr* running on the AS responds to the requests coming from a remote device's web server, to make the necessary conversion of data and files and to accomplish the synchronization between the local and the remote databases. The case study described in this paper uses this *hybrid* configuration: we can see the web application implementing the *medical process* distributed both on the device side, where a physician works with or without a connection with the server and signs the produced e-docs, and on the server side, where reside the HTML forms, the *Aplatform*, the AIDA tools and the SA servlet, all of them contributing in a correlated way at the generation and management of the e-docs. The choice to use this architecture has been determined essentially by two reasons. Firstly, the μ-web's main characteristic is to allow the applications operate both on-line and off-line. Additionally, this architecture has built-in the functionality needed to communicate and to exchange information between a PC server and a mobile device, so the development of an application running on the PDA addressed to update data on the server and client and to manage *Aplatform*'s service requests, turns out to be very easy. Second, it offers a user-friendly web interface, with which inexpert users can work without difficulties and where the data representation is completely transparent to the user. The SA is implemented as a set of Java servlets running on the AS. The whole process of creating the medical reports is divided in two distinct parts as it follows.

*Surgical Process.* This process involves not only the *Aplatform* used to generate e-MRs, but include also the set of activities involved to improve the efficiency of services offered to the patient and health office. All information about the surgery, e.g., reference to the previous medical examinations, will be stored in a DB on the AS. Part of this information will be downloaded on the specialist's mobile device in order to optimize the productivity of the specialist and the service usability. A *medical electronic register* is a database called *master-db* used for registering all the examinations both for both types of patients. The *master-db* is accessed via ODBC by a specific application whose role is to manage the reservation and registration phases besides other sanitary office's organizational operations. An authorized person inserts into the register data related to all specialist visits and instrumental examinations of the day, specifying the patient data, the examination type, the specialist that will visit this patient, the date of the visit and optionally the room where the examination will be performed. Each day the visits to be performed at this date are registered; this operation allows the creation of the list of patients to visit for each physician in the surgery. At start-up, after the authentication phase, each specialist downloads on his mobile terminal a list of patients to visit. For this purpose are used the terminal's web application, the μ–Web's API and data from the *master-db*. The patients' data is stored in a *light-db* residing on the specialist's terminal. For each patient are downloaded also from the *master-db* some additional data if existent (e.g., case sheet of previous admissions). For each visited patient the specialist runs next the *Medical Process*. The connection with the AS can be active only for the time of this download, so the specialist can use further the handheld PC offline during the whole examination.

*Medical Process.* During the visit, the data from the *light-db* is used to fill out a form containing patient information, (e.g., the surname). It is also possible to see the diagnosis of the previous examinations. When the visit ends up the specialist writes his diagnosis and the whole data is locally saved on the *light-db*. Both *light-db* and the *master-db* are relational databases. Afterwards the doctor has two possibilities. The first one is that he transparently uploads the data for the patient from the *light-db* on the *master-db*. For this purpose a function of the μ-web's API is called. The physician activates next the SA servlet that reads the data from the *master-db* and uses the AIDA API to create an unsigned e-MR containing the name, surname of the patient, the date of the visit, the type of visit performed and the corresponding diagnosis. This e-MR is sent back to the Jornada's browser where it can be further digitally signed using the AIDASign application. The second possibility is that the specialist continues with the next visit, if, for example, the connection with the remote server cannot be done. When the connection is active, the diagnoses saved on local storage are sent to the AS in order to update also the *master-db* by means of μ-web's API as it has been described above when the data related to a single patient had to be processed. Consequently, when needed, the physician sends requests to the IIS for the generation of *e-MR* from handheld PC. The Microsoft IIS, in conjunction with the servlet engine and the *Aplatform*, runs the requested servlet and generate an e-MR in XML format. The e-MR is sent back to the handheld PC's browser where it can be further digitally signed using the AIDASign. In both cases described above the produced signed *e-MR*s (stored as files) are transferred finally on the server by calling one function of µ-web server's API. Another servlet allows sending this file to the *Aplatform* to be stored in the *document directory*. Another function allow the physician to request the *ServerMgr* on the AS side to print out the *e-MR*. Finally, the signed e-MRs can be deleted from the mobile device in order to free space on the device. In this scenario there cannot be conflict between *master-db* and *light-db* because they are constantly synchronized. Only the physician that performs a visit accesses the patient data and at this moment all the additional information about the patient (i.e. results of other examinations) that is interesting for the diagnosis, must have been already inserted into the *master-db*.

## 5. CONCLUSION

In this paper we presented a solution for the management of electronic medical reports by making use of XML e-docs, embedded XML signatures and AIDA management platform for e-docs that can be easily adapted to various workflows, connectivity requirements and existent database back-ends. A handheld PC HP Jornada 720 that is provided with smart-card support for the generation of the digital signature is used as mobile signing terminal. The case study shows how this service can be used in a real medical environment where the users have intermittent network connectivity and consequently need to operate mostly off-line.

## ACKNOWLEDGEMENT

We would like to thank Prof. S. Pillon, the physician of Angiology division of S. Camillo-Forlanini Hospital for his continuous feedback and collaboration in the design of our application.

## REFERENCES

[1] http://aida.infonova.at/. Work funded by the European Commission under project AIDA (IST-1999-10497).

[2] Ayers D., H. Bergsten, M. Bogovich, J. Diamond. *Professional Java Server Programming*, Wrox Press Ltd., ISBN: 1861002777

[3] Berbecaru D., Lioy A., Marian M., 2001. A Framework for Secure Digital Administration, *Proceedings of the EuroWeb 2001- The Web in Public Administration*, Pisa, Italy, pp. 119-133.

[4] Bray T., Paoli J., Sperberg-McQueen C.M., *Extensible Markup Language (XML) 1.0, W3C Recommendation*, February 10, 1998. http://www.w3.org/TR/1998/REC-xml-19980210

[5] Eastlake D., Reagle J., Solo D., *XML-Signature Syntax and Processing*, RFC 3075, IETF, 2001

[6] Freudenthal M., Heiberg S., Willemson J., 2000. Personal Security Environment on Palm PDA, *Proceedings of Computer Security Applications Conference*, New Orleans, Louisiana, USA, pp. 366-372

[7] Scheibelhofer K., 2001. What You See Is What You Sign – Trustworthy Display of XML Documents for Signing and Verification, *Proceedings of CMS 2001 Conference*, Darmstadt, Germany, pp. 3-13

[8] Scheibelhofer K., 2001, What You See Is What You Sign, *Proceedings of the RSA Conference 2001 Europe (Online)*, Amsterdam, Netherlands

[9] Scheibelhofer K., 2000. Using OpenCard in Combination with the Java Cryptographic Architecture for Digital Signing, *Proceedings of the GEMPLUS Developer Conference*, Montpellier, France

[10] Scheibelhofer K., 2001. What You See Is What You Sign - Trustworthy Display of XML Documents for Signing and Verification, *Proceedings of CMS 2001 Conference*, Darmstadt, Germany, 21-22 May 2001, pp. 3-13

[11] Okamoto T., Tada M., Okamoto E., 1999. Extended Proxy Signatures for Smart Cards, *Proceedings of Information Security Conference*, vol. 1729 of LNCS, Kuala Lumpur, Malaysia, pp. 247-258

[12] Vedder K., Weikmann F., 1998. Smart Cards – Requirements, Properties, and Applications, *State of the Art in Applied Cryptography, vol. 1528 of LNCS*, pp. 307-331